\documentclass{IEEEcsmag}

\usepackage[colorlinks,urlcolor=blue,linkcolor=blue,citecolor=blue]{hyperref}

\expandafter\def\expandafter\UrlBreaks\expandafter{\UrlBreaks\do\/\do\*\do\-\do\~\do\'\do\"\do\-}
\usepackage{upmath}
\usepackage[dvipsnames]{xcolor}

\newcommand{\statbox}[2]{\noindent\raisebox{0.7\height}{\fcolorbox[rgb]{0,0,0}{#1}{\scriptsize~}}}

\usepackage{longfbox}

\newfboxstyle{tight2}{padding=2pt,margin=0pt,baseline-skip=false}%

\definecolor{ColT1}{HTML}{1f77b4}
\definecolor{ColT2}{HTML}{ff7f0e}
\definecolor{ColT3}{HTML}{2ca02c}
\definecolor{ColT4}{HTML}{d62728}
\definecolor{ColT5}{HTML}{9467bd}
\definecolor{ColT6}{HTML}{8c564b}
\definecolor{ColT7}{HTML}{e377c2}
\definecolor{ColT8}{HTML}{7f7f7f}
\definecolor{ColT9}{HTML}{bcbd22}
\definecolor{ColT10}{HTML}{17becf}

\newlength{\boxh}
\usepackage{upmath}

\jvol{XX}
\jnum{XX}
\paper{8}
\jmonth{March/April}
\jname{IEEE Computer Graphics and Applications}
\pubyear{2025}

\usepackage{tcolorbox}

\begin{document}

\sptitle{DEPARTMENT: DISSERTATION IMPACT}

\title{Visual Analytics for Explainable and Trustworthy Artificial Intelligence}

\author{Angelos Chatzimparmpas}
\affil{Utrecht University, Utrecht, 3584 CC, The Netherlands}

\markboth{DEPARTMENT: DISSERTATION IMPACT}{DEPARTMENT: DISSERTATION IMPACT}

\begin{abstract}\looseness-1Our society increasingly depends on intelligent systems to solve complex problems, ranging from recommender systems suggesting the next movie to watch to AI models assisting in medical diagnoses for hospitalized patients. With the iterative improvement of diagnostic accuracy and efficiency, AI holds significant potential to mitigate medical misdiagnoses by preventing numerous deaths and reducing an economic burden of approximately €450 billion annually. However, a key obstacle to AI adoption lies in the lack of transparency: many automated systems function as "black boxes," providing predictions without revealing the underlying processes. This opacity can hinder experts’ ability to trust and rely on AI systems. Visual analytics (VA) provides a compelling solution by combining AI models with interactive visualizations. These specialized charts and graphs empower users to incorporate their domain expertise to refine and improve the models, bridging the gap between AI and human understanding. In this work, we define, categorize, and explore how VA solutions can foster trust across the stages of a typical AI pipeline. We propose a design space for innovative visualizations and present an overview of our previously developed VA dashboards, which support critical tasks within the various pipeline stages, including data processing, feature engineering, hyperparameter tuning, understanding, debugging, refining, and comparing models.
\end{abstract}

\maketitle

\begin{tcolorbox}[colback=blue!5!white, colframe=blue!5!white, width=1.0\columnwidth]
Editor's note: 
\newline Angelos Chatzimparmpas received the 2024 EuroVis Best Ph.D. Dissertation Award.
\end{tcolorbox}

\vspace{1em}

\chapteri{T}he rapid improvement in hardware technology and the increased availability of computational resources has revamped the field of AI in recent years. These advancements did not go unnoticed by companies and other organizations wanting to use automated methods to enhance performance and provide better services to their clients. However, employing such AI solutions often comes with a trade-off: these models lack accountability for their predictions, and their decision making processes are opaque. This lack of transparency is problematic in high-stakes domains where experts need to understand the reasoning behind specific predictions to establish trust in AI systems.$^1$ For example, in healthcare, diagnostic errors contribute significantly to preventable harm, with misdiagnoses accounting for €446 billion in wasted costs in the EU (as reported by the Symptoma EU project) and \$750 billion in the US. Annually, around 795,000 Americans die or suffer permanent disability due to misdiagnoses of severe illnesses like cancer, one of the "big three" most frequently misdiagnosed conditions.$^2$ While AI holds the potential to revolutionize medical practice, its successful integration into application domains like digital pathology remains limited.$^3$

\subsection{Artificial Intelligence Research}
Initial efforts of AI researchers primarily targeted methods for collecting and labeling large datasets to serve as inputs for progressively more sophisticated AI models. Within the AI community, most studies were centered around achieving better predictive performance. After significant advancements in hyperparameter optimization and algorithmic development, there appears to be a renewed focus on data-centric AI.$^4$ Intrinsic algorithmic interpretability and post-hoc explainability have also become indispensable components of modern AI pipelines and are now more critical than ever as sanity checks and assessment tools of an AI model's overall knowledge.

While the achievements from the AI domain are remarkable, particularly with fully automated computational methods tailored for specific tasks, much of the research has fallen short of addressing human needs in this area. Moreover, disciplines like psychology and cognitive science have long explored concepts such as trust, interpretability, and explainability. The information visualization (InfoVis) and visual analytics (VA) communities also intersect here, offering techniques that leverage human perception and cognitive abilities.

\begin{figure*}
\centerline{\includegraphics[width=36pc]{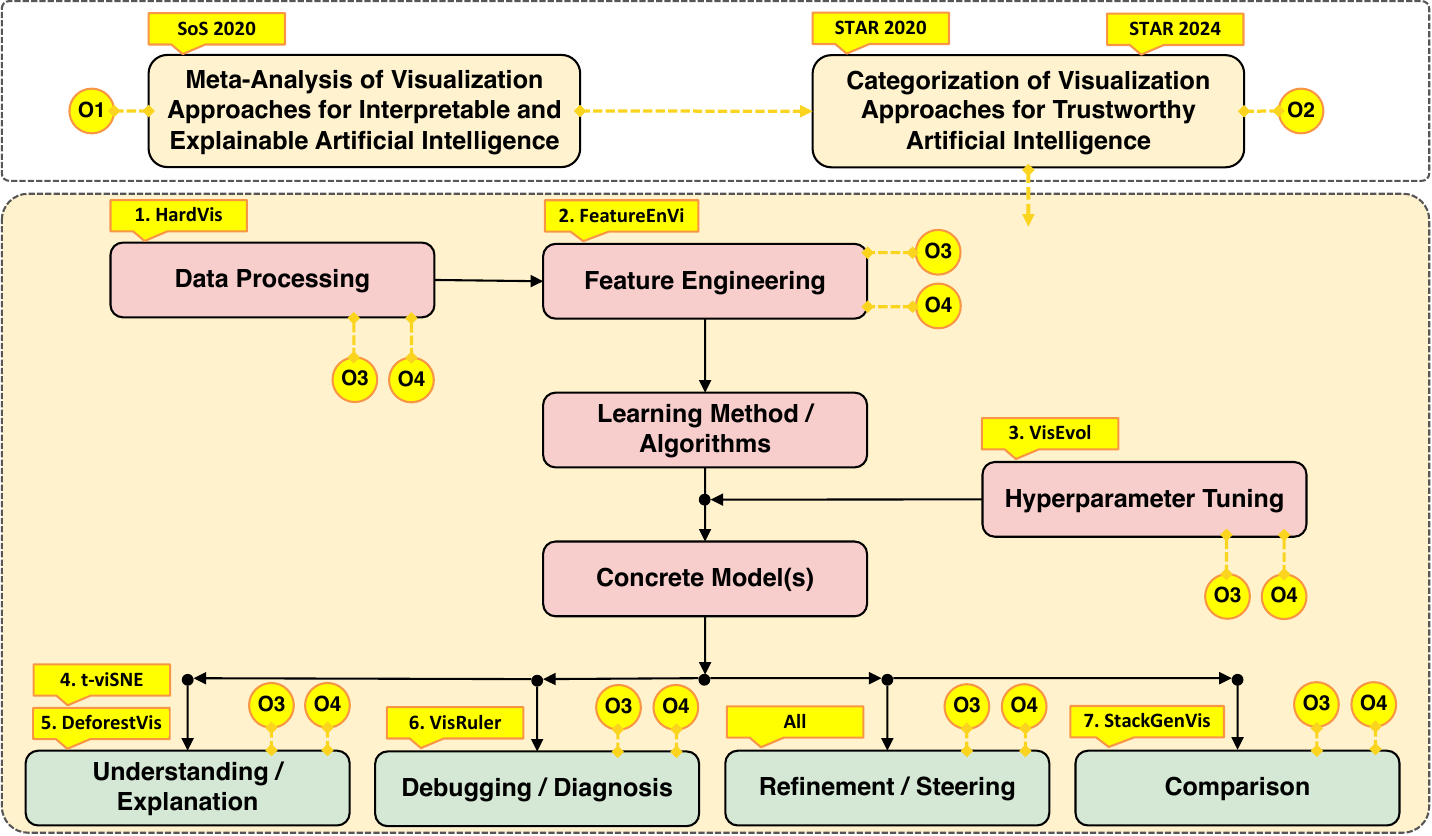}}
\caption{Overview of the main results of our research. The first block at the top presents three consensus reviews, encoding the generated knowledge in light yellow. This knowledge is used to identify research gaps and contribute to the field with our own works (1--7), as shown in the second block. The red boxes present a typical AI pipeline with four trust components in green. Lastly, the vivid yellow nodes refer to the research objectives (O1--O4) associated with the related results.}
\label{fig:outline}
\end{figure*}

\subsection{Visual Analytics Research}
InfoVis and VA are subfields of the broader area of human-computer interaction (HCI) research. The former focuses on leveraging human cognitive abilities through the use of interactive, computer-supported data visualizations,$^5$ while the latter centers on using interactive visual interfaces to facilitate analytical reasoning processes.$^6$ In our context, a typical VA approach integrates methods from both the InfoVis/VA and AI domains, aiming to combine computational analyses with interactive visualizations. By doing so, the VA dashboard collaborates with the human analyst to interpret source data, comprehend models and predictions, and acquire knowledge, thereby completing the sensemaking loop. Consequently, a human-centered strategy for explainable AI (XAI) supports analysts in understanding data when fully automated methods either fall short or are unsuitable.

The primary issue in this research domain is that most methodologies either emphasize understanding AI models while neglecting the initial phases of the AI pipeline, such as data processing and feature engineering, or focus exclusively on generating explanations for complex models like neural networks (NNs). However, deep learning may not be particularly effective for tabular data comprising meaningful features,$^7$ especially when compared to ensemble learning methods that are often more powerful. Therefore, the main goals (Gs) of this work are to: (G1) define the concept of trust in the context of AI, (G2) examine and classify existing research works, (G3) propose integrated workflows with interactive visualization (VIS) for XAI across all stages, (G4) create model-agnostic VA dashboards aligned with the ensemble learning paradigm, and (G5) implement a model-specific dimensionality reduction solution that enhances the robustness of the results.

\section{OBJECTIVES}

To fulfill these goals, we define the following four research objectives (Os) that are also visible in Fig.~\ref{fig:outline}:

O1 involves conducting a meta-analysis of the current state of interactive visualization use for XAI (G1). This is achieved by manually collecting survey papers and analyzing their perspectives on individual peer-reviewed papers, summarizing key insights, and creating a guide to help researchers navigate this vast domain. This step establishes fundamental knowledge of the field and highlights research opportunities. It confirms the growing trend of using visualization to interpret and explain AI models in recent years, showing its potential to enhance trust in such models.

O2 focuses on a comprehensive literature review of how interactive visualization contributes to building trust in AI models (G2). We establish a design space by providing background on trustworthiness, reviewing relevant literature, categorizing visualization techniques extensively, and identifying future research directions. Our work offers a categorization of trust across various stages of interactive AI (cf. Table~\ref{tab:categorization}).

O3 targets the development of unified workflows addressing key research questions at different stages of the AI pipeline (G3). For example, during the initial stage, we enable users to identify and sample training instances---both easy and challenging to classify---from all classes rather than uniformly applying under or oversampling. In the next stages, we introduce methods like data space slicing for feature exploration and model generation through evolutionary hyperparameter optimization processes. Users are also supported in evaluating global accuracy and local quality with projection-based visualizations, facilitating better understanding and debugging of ensemble models. Lastly, we aid users in comparing models to streamline the creation of model stacks by eliminating underperforming or overly complex models.

O4 focuses on translating analytical tasks into specific design goals for building concrete VA dashboards (G4 and G5). We contribute by designing and implementing multiple interactive VA techniques. These proof-of-concept dashboards are validated using both quantitative and qualitative methods by AI and domain experts. By practically implementing theoretical ideas, our dashboards aim to cover the defined design space.

\begin{table*}[t!]
\centering
	\caption{AI visualization techniques categorization in the 2024 STAR$^1$$^0$ (including comparison with the 2020 STAR data).}
	\setlength{\tabcolsep}{2pt}
	\renewcommand{\arraystretch}{1}
	\setlength\doublerulesep{2mm} 
	\tiny
	\begin{minipage}[t]{0.36\textwidth}
	\begin{tabular}[t]{|l|r|c|}
	\hline \textbf{Domain} & \textbf{542}~\statbox{0.00,0.50,0.00}{100\%} & \\
	\hline Biology & 49~\statbox{0.91,0.96,0.91}{9\%} & -5\%~$\downarrow$\\ 
	\hline Business & 87~\statbox{0.84,0.92,0.84}{16\%} & +6\%~$\uparrow$\\
	\hline Computer Vision & 180~\statbox{0.67,0.84,0.67}{33\%} & +3\%~$\uparrow$\\
	\hline Computers & 9~\statbox{0.98,0.99,0.98}{2\%} & -1\%~$\downarrow$\\
	\hline Health & 90~\statbox{0.84,0.92,0.84}{17\%} & +2\%~$\uparrow$\\
	\hline Humanities & 81~\statbox{0.85,0.93,0.85}{15\%} & -6\%~$\downarrow$\\
	\hline Nutrition & 16~\statbox{0.97,0.98,0.97}{3\%} & -1\%~$\downarrow$\\
	\hline Simulation & 19~\statbox{0.96,0.98,0.96}{4\%} & 0\%~~-\\
	\hline Social / Socioeconomic & 58~\statbox{0.89,0.95,0.89}{11\%} & 0\%~~-\\
	\hline Other & 230~\statbox{0.58,0.79,0.58}{42\%} & -5\%~$\downarrow$\\
	\hline 
	\hline \textbf{Target Variable} & \textbf{542}~\statbox{0.00,0.50,0.00}{100\%} & \\ 
	\hline Binary (categorical) & 85~\statbox{0.84,0.92,0.84}{16\%} & -4\%~$\downarrow$\\
	\hline Multi-class (categorical) & 297~\statbox{0.45,0.73,0.45}{55\%} & -9\%~$\downarrow\downarrow$\\
	\hline Multi-label (categorical) & 21~\statbox{0.96,0.98,0.96}{4\%} & -1\%~$\downarrow$\\
	\hline Continuous (regression problems) & 60~\statbox{0.88,0.94,0.88}{12\%} & 0\%~~-\\
	\hline Other & 147~\statbox{0.73,0.87,0.73}{27\%} & +8\%~$\uparrow$\\
	\hline 
	\hline \textbf{AI Methods} & \textbf{542}~\statbox{0.00,0.50,0.00}{100\%} & \\ 
	\hline Convolutional Neural Network (CNN) & 72~\statbox{0.87,0.93,0.87}{13\%} & 0\%~~-\\
	\hline Deep Convolutional Network (DCN) & 12~\statbox{0.98,0.99,0.98}{2\%} & -2\%~$\downarrow$\\
	\hline Deep Feed Forward (DFF) & 11~\statbox{0.98,0.99,0.98}{2\%} & -3\%~$\downarrow$\\
	\hline Deep Neural Network (DNN) & 55~\statbox{0.90,0.95,0.90}{10\%} & 0\%~~-\\
	\hline Deep Q-Network (DQN)& 11~\statbox{0.98,0.99,0.98}{2\%} & -3\%~$\downarrow$\\
	\hline Generative Adversarial Network (GAN) & 13~\statbox{0.98,0.99,0.98}{2\%} & -3\%~$\downarrow$\\
	\hline Long Short-Term Memory (LSTM) & 26~\statbox{0.95,0.98,0.95}{5\%} & -2\%~$\downarrow$\\
	\hline Recurrent Neural Network (RNN) & 31~\statbox{0.94,0.97,0.94}{6\%} & -3\%~$\downarrow$\\
	\hline Variational Auto-Encoder (VAE) & 22~\statbox{0.96,0.98,0.96}{4\%} & -3\%~$\downarrow$\\
	\hline Other (DL methods) & 92~\statbox{0.84,0.92,0.84}{17\%} & +6\%~$\uparrow$\\
	\hline Linear (DR) & 81~\statbox{0.85,0.93,0.85}{15\%} & -14\%~$\downarrow\downarrow$\\
	\hline Non-linear (DR) & 94~\statbox{0.83,0.91,0.83}{17\%} & -9\%~$\downarrow\downarrow$\\
	\hline Bagging (ensemble learning) & 74~\statbox{0.86,0.93,0.86}{14\%} & 0\%~~-\\
	\hline Boosting (ensemble learning) & 30~\statbox{0.94,0.97,0.94}{6\%} & 0\%~~-\\
	\hline Stacking (ensemble learning) & 10~\statbox{0.98,0.99,0.98}{2\%} & -1\%~$\downarrow$\\
	\hline Other (generic)& 233~\statbox{0.57,0.78,0.57}{43\%} & -6\%~$\downarrow$\\
	\hline 
	\hline \textbf{AI Types} & \textbf{500}~\statbox{0.08,0.54,0.08}{92\%} & \\ 
	\hline Classification (supervised)& 288~\statbox{0.47,0.74,0.47}{53\%} & -3\%~$\downarrow$\\
	\hline Regression (supervised) & 55~\statbox{0.90,0.95,0.90}{10\%} & 0\%~~-\\
	\hline Other (supervised)& 32~\statbox{0.95,0.97,0.95}{6\%} & +2\%~$\uparrow$\\
	\hline Association (unsupervised) & 14~\statbox{0.97,0.99,0.97}{3\%} & 0\%~~-\\
	\hline Clustering (unsupervised)& 69~\statbox{0.87,0.94,0.87}{13\%} & -8\%~$\downarrow$\\
	\hline Dimensionality Reduction (unsupervised)& 104~\statbox{0.81,0.90,0.81}{19\%} & -14\%~$\downarrow\downarrow$\\
	\hline Classification (semi-supervised)& 40~\statbox{0.93,0.96,0.93}{7\%} & 0\%~~-\\
	\hline Clustering (semi-supervised)& 16~\statbox{0.97,0.98,0.97}{3\%} & 0\%~~-\\
	\hline Classification (reinforcement)& 4~\statbox{1.00,1.00,1.00}{1\%} & 0\%~~-\\
	\hline Control (reinforcement)& 11~\statbox{0.98,0.99,0.98}{2\%} & 0\%~~-\\
	\hline	
	\end{tabular} 
	\bigbreak
	Color Legend: \thinspace
	\statbox{1.00,1.00,1.00}{0\%} {\strut 0 papers} \thinspace
	\statbox{0.50,0.75,0.50}{50\%} {\strut 271 papers} \thinspace
	\statbox{0.00,0.50,0.00}{100\%} {\strut 542 papers}
	\end{minipage}
	\hspace{1mm}
	\begin{minipage}[t]{0.33\textwidth}
	\begin{tabular}[t]{|l|r|c|}
	\hline \textbf{AI Processing Phase} & \textbf{542}~\statbox{0.00,0.50,0.00}{100\%} & \\ 
	\hline Pre-processing / Input & 120~\statbox{0.78,0.89,0.78}{22\%} & +4\%~$\uparrow$\\
	\hline In-processing / Model & 156~\statbox{0.71,0.85,0.71}{29\%} & +6\%~$\uparrow$\\
	\hline Post-processing / Output & 363~\statbox{0.33,0.67,0.33}{67\%} & -14\%~$\downarrow\downarrow$\\
	\hline
	\hline \textbf{Treatment Method} & \textbf{542}~\statbox{0.00,0.50,0.00}{100\%} & \\ 
	\hline Model-agnostic / Black Box & 395~\statbox{0.27,0.64,0.27}{73\%} & +1\%~$\uparrow$\\
	\hline Model-specific / White Box & 165~\statbox{0.69,0.85,0.69}{30\%} & -5\%~$\downarrow$\\
	\hline 
	\hline \textbf{Dimensionality} & \textbf{542}~\statbox{0.00,0.50,0.00}{100\%} & \\ 
	\hline 2D & 538~\statbox{0.01,0.51,0.01}{99\%} & +1\%~$\uparrow$\\
	\hline 3D & 9~\statbox{0.98,0.99,0.98}{2\%} & -1\%~$\downarrow$\\
	\hline
	\hline \textbf{Visual Aspects} & \textbf{542}~\statbox{0.00,0.50,0.00}{100\%} & \\ 
	\hline Computed & 497~\statbox{0.08,0.55,0.08}{92\%} & -6\%~$\downarrow$\\
	\hline Mapped & 273~\statbox{0.50,0.75,0.50}{50\%} & -5\%~$\downarrow$\\
	\hline 
	\hline \textbf{Visual Granularity} & \textbf{542}~\statbox{0.00,0.50,0.00}{100\%} & \\ 
	\hline Aggregated Information & 468~\statbox{0.14,0.57,0.14}{86\%} & -6\%~$\downarrow$\\
	\hline Instance-based / Individual & 377~\statbox{0.31,0.65,0.31}{70\%} & -3\%~$\downarrow$\\
	\hline 
	\hline \textbf{Visual Representation} & \textbf{542}~\statbox{0.00,0.50,0.00}{100\%} & \\ 
	\hline Bar Charts & 243~\statbox{0.55,0.77,0.55}{45\%} & +4\%~$\uparrow$\\
	\hline Box Plots & 39~\statbox{0.93,0.96,0.93}{7\%} & +1\%~$\uparrow$\\
	\hline Matrix & 153~\statbox{0.72,0.86,0.72}{28\%} & +3\%~$\uparrow$\\
	\hline Glyphs / Icons / Thumbnails & 131~\statbox{0.76,0.88,0.76}{24\%} & -8\%~$\downarrow$\\
	\hline Grid-based Approaches & 59~\statbox{0.89,0.95,0.89}{11\%} & +1\%~$\uparrow$\\
	\hline Heatmaps & 149~\statbox{0.73,0.86,0.73}{27\%} & +4\%~$\uparrow$\\
	\hline Histograms & 174~\statbox{0.68,0.84,0.68}{32\%} & +4\%~$\uparrow$\\
	\hline Icicle Plots & 9~\statbox{0.98,0.99,0.98}{2\%} & -1\%~$\downarrow$\\
	\hline Line Charts & 154~\statbox{0.72,0.86,0.72}{28\%} & 0\%~~-\\
	\hline Node-link Diagrams & 128~\statbox{0.76,0.88,0.76}{24\%} & 0\%~~-\\
	\hline Parallel Coordinates Plots (PCPs) & 73~\statbox{0.87,0.93,0.87}{13\%} & -3\%~$\downarrow$\\
	\hline Pixel-based Approaches & 20~\statbox{0.96,0.98,0.96}{4\%} & 0\%~~-\\
	\hline Radial Layouts & 90~\statbox{0.84,0.92,0.84}{17\%} & +6\%~$\uparrow$\\
	\hline Scatterplot Matrices (SPLOMs) & 26~\statbox{0.95,0.97,0.95}{5\%} & -4\%~$\downarrow$\\
	\hline Scatterplot / Projections & 294~\statbox{0.46,0.73,0.46}{54\%} & -4\%~$\downarrow$\\
	\hline Similarity Layouts & 190~\statbox{0.65,0.82,0.65}{35\%} & +21\%~$\uparrow\uparrow\uparrow$\\
	\hline Tables / Lists & 201~\statbox{0.63,0.82,0.63}{37\%} & -6\%~$\downarrow$\\
	\hline Treemaps & 13~\statbox{0.98,0.99,0.98}{2\%} & -1\%~$\downarrow$\\
	\hline Other & 207~\statbox{0.62,0.81,0.62}{38\%} & +8\%~$\uparrow$\\
	\hline 
	\hline \textbf{Interaction Technique} & \textbf{525}~\statbox{0.03,0.52,0.03}{97\%} & \\ 
	\hline Select & 491~\statbox{0.09,0.55,0.09}{91\%} & +9\%~$\uparrow\uparrow$\\
	\hline Explore / Browse & 445~\statbox{0.18,0.59,0.18}{82\%} & -3\%~$\downarrow$\\
	\hline Reconfigure & 220~\statbox{0.59,0.80,0.59}{41\%} & +4\%~$\uparrow$\\
	\hline Encode & 294~\statbox{0.46,0.73,0.46}{54\%} & -2\%~$\downarrow$\\
	\hline Filter / Query & 314~\statbox{0.42,0.71,0.42}{58\%} & +1\%~$\uparrow$\\
	\hline Abstract / Elaborate & 354~\statbox{0.35,0.67,0.35}{65\%} & -24\%~$\downarrow\downarrow\downarrow$\\
	\hline Connect & 395~\statbox{0.27,0.64,0.27}{73\%} & +9\%~$\uparrow\uparrow$\\
	\hline Guide / Sheperd & 169~\statbox{0.69,0.84,0.69}{31\%} & +7\%~$\uparrow$\\
	\hline Verbalize & 31~\statbox{0.94,0.97,0.94}{6\%} & +1\%~$\uparrow$\\
	\hline 
	\end{tabular} 
	\end{minipage}
	\hspace{1mm}
	\begin{minipage}[t]{0.28\textwidth}
	\begin{tabular}[t]{|l|r|c|}
	\hline \textbf{Visual Variable} & \textbf{542}~\statbox{0.00,0.50,0.00}{100\%} & \\ 
	\hline Color & 540~\statbox{0.00,0.50,0.00}{100\%} & +2\%~$\uparrow$\\
	\hline Opacity & 294~\statbox{0.46,0.73,0.46}{54\%} & +12\%~$\uparrow\uparrow$\\
	\hline Position / Orientation & 83~\statbox{0.85,0.93,0.85}{15\%} & -14\%~$\downarrow\downarrow$\\
	\hline Shape & 89~\statbox{0.84,0.92,0.84}{16\%} & -3\%~$\downarrow$\\
	\hline Size & 164~\statbox{0.70,0.85,0.70}{30\%} & -4\%~$\downarrow$\\
	\hline Texture & 30~\statbox{0.94,0.97,0.94}{6\%} & -3\%~$\downarrow$\\
	\hline 
	\hline \textbf{Evaluation} & \textbf{542}~\statbox{0.00,0.50,0.00}{100\%} & \\ 
	\hline Standard & 132~\statbox{0.76,0.88,0.76}{24\%} & +5\%~$\uparrow$\\
	\hline Comparative & 27~\statbox{0.95,0.98,0.95}{5\%} & -1\%~$\downarrow$\\
	\hline Before / During Development & 140~\statbox{0.74,0.87,0.74}{26\%} & +8\%~$\uparrow$\\
	\hline After Development & 175~\statbox{0.68,0.84,0.68}{32\%} & +8\%~$\uparrow$\\
	\hline Not Evaluated & 196~\statbox{0.64,0.82,0.64}{36\%} & -15\%~$\downarrow\downarrow$\\
	\hline 	
	\hline \textbf{Trust Levels (TL) 1--5} & \textbf{542}~\statbox{0.00,0.50,0.00}{100\%} & \\ 
	\hline Source Reliability & 43~\statbox{0.92,0.96,0.92}{8\%} & +2\%~$\uparrow$\\
	\hline Transparent Collection Process & 21~\statbox{0.96,0.98,0.96}{4\%} & +1\%~$\uparrow$\\
	\noalign{
	\global\dimen1\arrayrulewidth
	\global\arrayrulewidth1pt
	}
	\hline
	\noalign{
	\global\arrayrulewidth\dimen1 
	}Uncertainty Awareness & 84~\statbox{0.85,0.93,0.85}{15\%} & 0\%~~-\\
	\hline Equality / Data Bias & 57~\statbox{0.89,0.95,0.89}{11\%} & +3\%~$\uparrow$\\
	\hline Comparison (of Structures) & 271~\statbox{0.50,0.75,0.50}{50\%} & +7\%~$\uparrow$\\
	\hline Guidance / Recommendations & 142~\statbox{0.74,0.87,0.74}{26\%} & +3\%~$\uparrow$\\
	\hline Outlier Detection & 174~\statbox{0.68,0.84,0.68}{32\%} & -2\%~$\downarrow$\\
	\noalign{
	\global\dimen1\arrayrulewidth
	\global\arrayrulewidth1pt
	}
	\hline
	\noalign{
	\global\arrayrulewidth\dimen1 
	}Familiarity & 24~\statbox{0.96,0.98,0.96}{4\%} & +2\%~$\uparrow$\\
	\hline Understanding / Explanation & 314~\statbox{0.42,0.71,0.42}{58\%} & +10\%~$\uparrow\uparrow$\\
	\hline Debugging / Diagnosis & 129~\statbox{0.76,0.88,0.76}{24\%} & -3\%~$\downarrow$\\
	\hline Refinement / Steering & 164~\statbox{0.70,0.85,0.70}{30\%} & -5\%~$\downarrow$\\
	\hline Comparison & 114~\statbox{0.79,0.89,0.79}{21\%} & -10\%~$\downarrow\downarrow$\\
	\hline Knowledgeability & 47~\statbox{0.91,0.96,0.91}{9\%} & +4\%~$\uparrow$\\
	\hline Fairness & 29~\statbox{0.95,0.97,0.95}{5\%} & +2\%~$\uparrow$\\
	\noalign{
	\global\dimen1\arrayrulewidth
	\global\arrayrulewidth1pt
	}
	\hline
	\noalign{
	\global\arrayrulewidth\dimen1 
	}Experience & 31~\statbox{0.94,0.97,0.94}{6\%} & +2\%~$\uparrow$\\
	\hline In Situ Comparison & 135~\statbox{0.75,0.87,0.75}{25\%} & -2\%~$\downarrow$\\
	\hline Performance & 338~\statbox{0.38,0.69,0.38}{62\%} & +8\%~$\uparrow$\\
	\hline What-if Hypotheses & 87~\statbox{0.84,0.92,0.84}{16\%} & -4\%~$\downarrow$\\
	\hline Model Bias & 58~\statbox{0.89,0.95,0.89}{11\%} & -1\%~$\downarrow$\\
	\hline Model Variance & 34~\statbox{0.94,0.97,0.94}{6\%} & -2\%~$\downarrow$\\
	\noalign{
	\global\dimen1\arrayrulewidth
	\global\arrayrulewidth1pt
	}
	\hline
	\noalign{
	\global\arrayrulewidth\dimen1 
	}Agreement of Colleagues & 21~\statbox{0.96,0.98,0.96}{4\%} & -1\%~$\downarrow$\\
	\hline Visualization Evaluation & 301~\statbox{0.44,0.72,0.44}{56\%} & +12\%~$\uparrow\uparrow$\\
	\hline Metrics Validation / Results & 413~\statbox{0.24,0.62,0.24}{76\%} & +11\%~$\uparrow\uparrow$\\
	\hline User Bias & 24~\statbox{0.96,0.98,0.96}{4\%} & 0\%~~-\\
	\hline 
	\hline \textbf{Target Group} & \textbf{542}~\statbox{0.00,0.50,0.00}{100\%} & \\ 
	\hline Beginners & 104~\statbox{0.82,0.91,0.82}{19\%} & -2\%~$\downarrow$\\
	\hline Practitioners / Domain Experts & 424~\statbox{0.22,0.61,0.22}{78\%} & -3\%~$\downarrow$\\
	\hline Developers & 113~\statbox{0.79,0.90,0.79}{21\%} & +3\%~$\uparrow$\\
	\hline AI Experts & 204~\statbox{0.62,0.81,0.62}{38\%} & +1\%~$\uparrow$\\
	\hline 	
	\end{tabular} 
 	\bigbreak
	Symbol Legend: 
    \\ \thinspace
        +/-[1 -- 8]\% $\uparrow$/$\downarrow$ \thinspace
        \\ +/-[9 -- 16]\% $\uparrow\uparrow$/$\downarrow\downarrow$ \thinspace
        \\ +/-[17 -- 24]\% $\uparrow\uparrow\uparrow$/$\downarrow\downarrow\downarrow$
	\end{minipage} 

	\medskip
	\scriptsize \emph{Note:} Each row shows the total count of techniques per category (with heatmap-style icons), and the \% difference compared to the 2020 STAR.
	\label{tab:categorization}
\end{table*}

\section{THE STATE OF THE ART} \label{sec:STAR}
To lay the foundation for this research, we draw on findings from a survey of surveys (SoS$^8$) and two literature reviews (STARs$^9$$^,$$^1$$^0$) that we conducted.

To explore the existing work in the design space of VIS for XAI, we begin by examining a set of 18 survey papers. The new insights obtained during this process help us pinpoint gaps in the literature that may appeal to early-stage and senior visualization researchers as well as AI experts.

\subsection{Online Training Processes}
A core future challenge is related to real-time online training processes (11 out of 18). This challenge arises because current methods mainly focus on analyzing the outcomes of AI techniques, neglecting the potential for an interactive training process that could enhance trust in the results. For example, consider an expert training an NN, a process that may take hours to complete. While the expert could use an output visualization to interpret the final results, a VA dashboard that supports the online training process could provide preliminary insights and thereby enable the expert to guide the training more effectively.

\subsection{Enhancing Trust in Artificial Intelligence}
Another key challenge is how to build trust in AI algorithms and models (10 out of 18).$^1$$^1$ Achieving this requires integrating interactive visualizations with traditional quantitative metrics into VA dashboards, thereby creating an "ecosystem" that allows expert users to oversee and manage AI models at various levels. For example, a VA dashboard might display how standard validation metrics (e.g., accuracy, precision, or recall) of classifiers evolve during each step of an AI algorithm's execution, enabling experts to detect deviations in the model's behavior. Thus, experts can intervene by inserting their knowledge into the process. In essence, VA dashboards should support timely interventions when needed.

To first define what trust in AI is, we propose a more detailed, multi-level model of trust in AI that consists of five trust levels (TLs): the raw data (TL1: source reliability and transparent collection process), the processed data (TL2: uncertainty awareness, equality/data bias, comparison of structures, guidance/recommendation, and outlier detection), the algorithm/learning method (TL3: familiarity, understanding/ explanation, debugging/diagnosis, refinement/steering, comparison, knowledgeability, and fairness), the concrete model(s) for a particular task (TL4: experience, in situ comparison, performance, what-if hypotheses, model bias, and model variance), and the evaluation and the subjective users’ expectations (TL5: agreement of colleagues, visualization evaluation, metrics validation/results, and user bias).

We then select 542 peer-reviewed papers that introduce a large variety of visualization techniques for increasing trust in AI models and their results. Based on them, we propose a fine-grained categorization comprising eight high-level design space aspects partitioned into 18 category groups that, on their part, contain 119 categories in total (see Table~\ref{tab:categorization}).

Our categorization served as a foundation for the design space of VA approaches for AI, which we introduce in the subsequent section.

\section{VA WORKFLOWS \& DASHBOARDS}

\subsection{Data Processing}

Despite significant advancements in AI, handling imbalanced data remains a challenge in many real-world scenarios. Among various approaches, sampling algorithms are considered effective solutions. However, many studies emphasize the major role of instance hardness, that is, the challenge of managing noisy or difficult-to-classify instances that leads to misclassification and degraded performance.

\begin{figure*}
\centerline{\includegraphics[width=36pc]{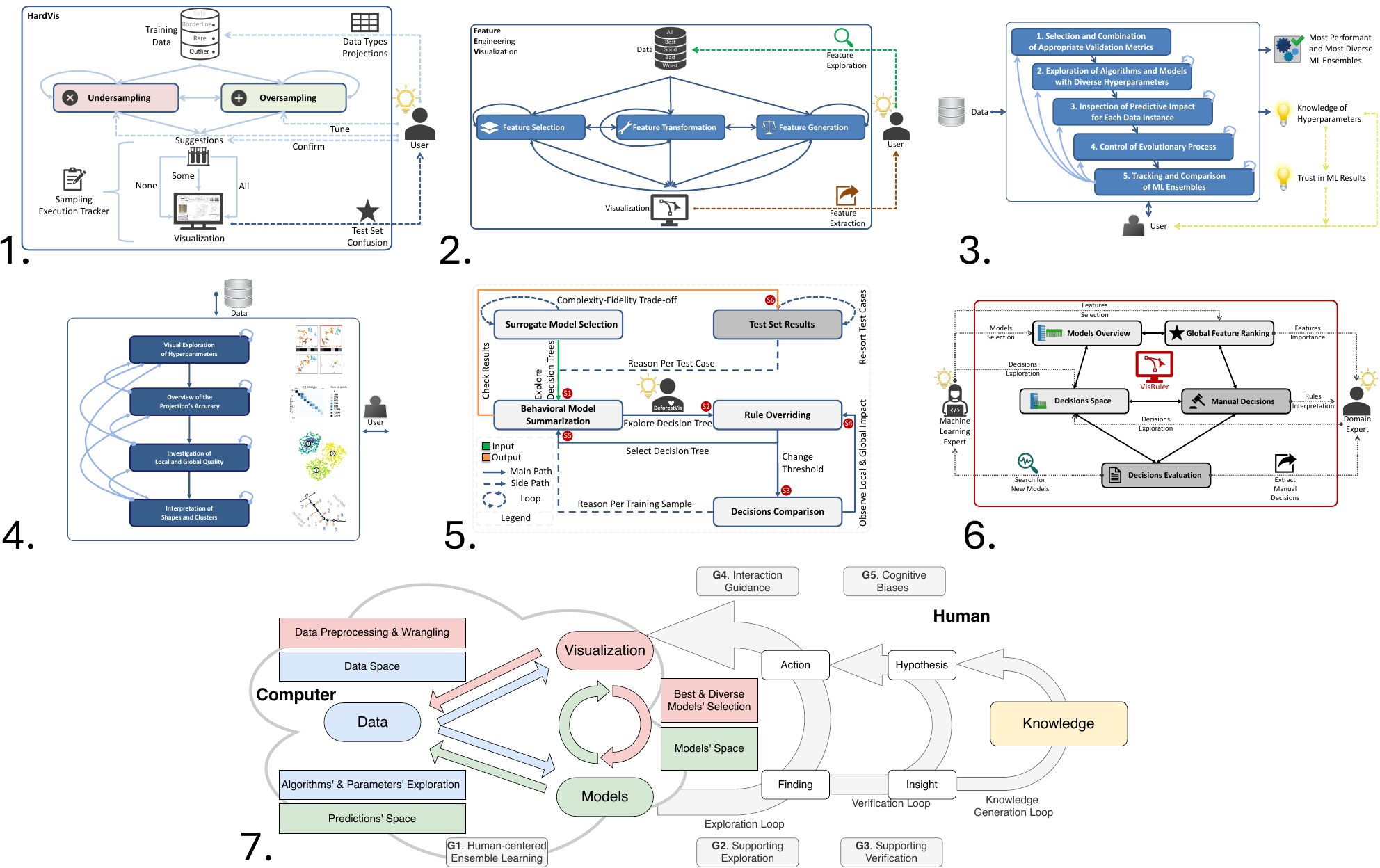}}
\caption{We designed coherent workflows for our seven technical contributions, covering each stage of the AI pipeline.}
\label{fig:workflow}
\end{figure*}

The proposed workflow begins with splitting the training data into four types based on the user's visual inspection of 9 alternative projections (see Fig.~\ref{fig:workflow}(1)). Data undergoes iterative undersampling or oversampling, with suggestions provided continuously and user confirmation sought after exploratory visual analysis.

HardVis$^1$$^2$ is a VA dashboard designed to address instance hardness by employing highly-configurable undersampling and oversampling methods (cf. Fig.~\ref{fig:system}(1)). The dashboard provides multiple coordinated views, enabling users to determine the optimal distribution of data types iteratively, selectively remove safe-to-discard samples, and interactively oversample others as needed. It also supports the analysis of algorithmic recommendations by using diverse visual cues to validate the safety of proposed removals or additions. This VA approach makes the entire process transparent by being effective in tackling instance hardness and class imbalance problems.

A limitation identified by AI experts is the scalability due to inadequate support for feature engineering.

\subsection{Feature Engineering}

Feature engineering can significantly benefit AI by enhancing predictive performance, minimizing computational costs, reducing unnecessary noise, and improving the transparency of decisions made during training. However, although various VA dashboards exist to monitor and control the different stages of the AI life cycle (e.g., those concerning data and algorithms), support for feature engineering is yet insufficient.

The proposed workflow starts by dividing the dataset based on the prediction probabilities of individual samples (see Fig.~\ref{fig:workflow}(2)). The data then undergoes three distinct feature engineering processes (selection, transformation, and generation) that are performed iteratively, with the user in control of the process.

FeatureEnVi$^1$$^3$ is a VA dashboard designed to support users in engineering features through stepwise feature selection and semi-automatic feature extraction methods (cf Fig.~\ref{fig:system}(2)). The dashboard employs multiple coordinated views to assist users in selecting, transforming, and creating new features as part of a highly iterative workflow. By examining the impact of features using various statistical metrics and automated feature selection techniques, users can enhance predictive performance, minimize computational resource requirements, and shorten training time. This VA approach increases adaptability and transparency by improving the feature engineering process.

A limitation is the lack of experimentation with alternative hyperparameter tuning techniques.

\subsection{Hyperparameter Tuning}

During the training phase of AI models, multiple hyperparameters typically need to be configured. This task is computationally demanding and involves a comprehensive search to identify the optimal hyperparameter set for a specific problem. The difficulty increases due to the fact that each hyperparameter in an AI algorithm is often interdependent with others, and modifying one may lead to unpredictable effects on the rest. Evolutionary hyperparameter optimization offers a promising approach to tackle these challenges. This strategy retains high-performing models while enhancing others through crossover and mutation techniques inspired by genetic algorithms.

The proposed workflow enables users to build effective and diverse AI ensembles while gaining insights into the hyperparameter selection process facilitated by evolutionary optimization (see Fig.~\ref{fig:workflow}(3)). Users can control the process by specifying the number of models involved in crossover and mutation operations for various algorithms.

\begin{figure*}
\centerline{\includegraphics[width=36pc]{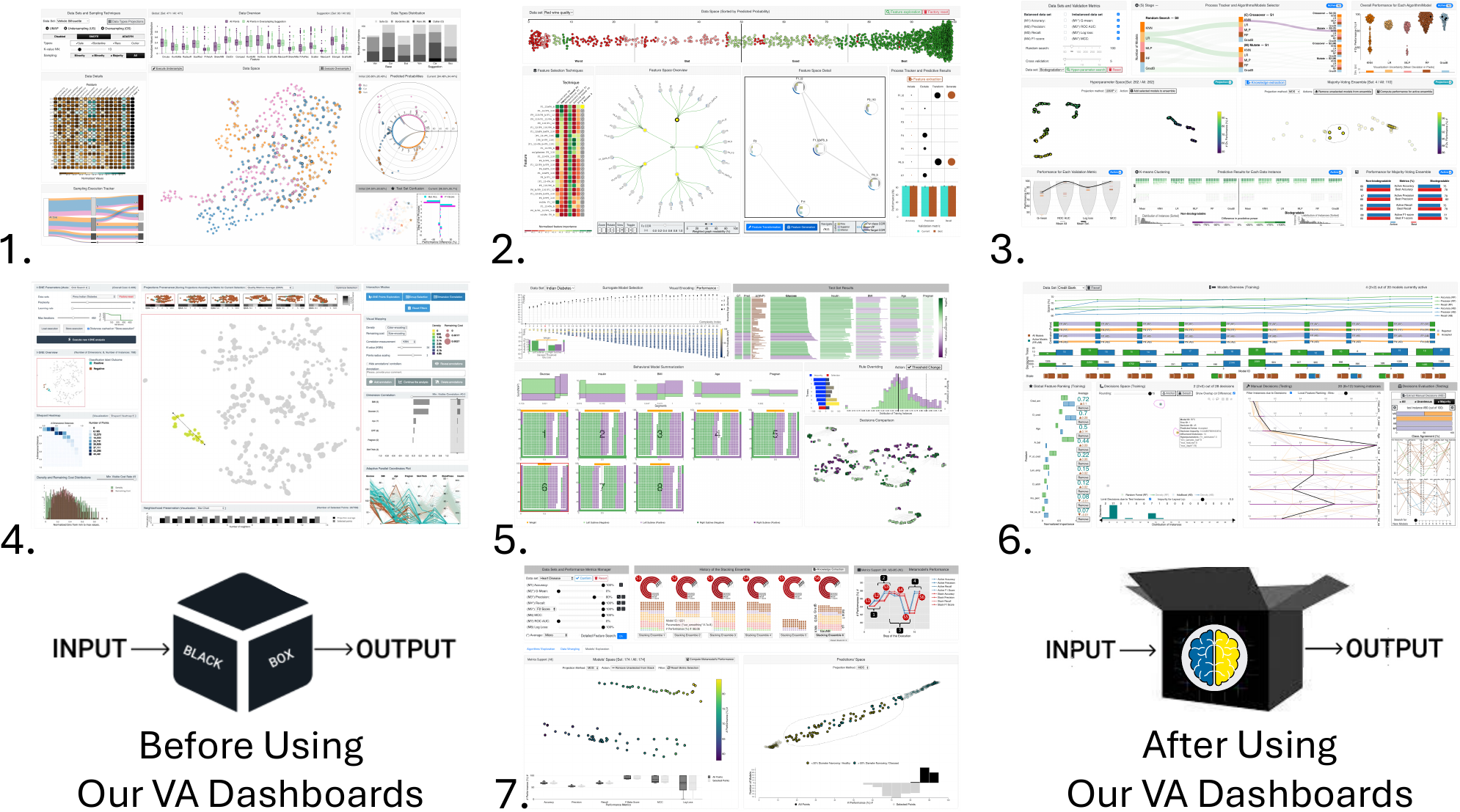}}
\caption{We developed seven visual analytics dashboards to address challenges at every stage of the AI pipeline effectively.}
\label{fig:system}
\end{figure*}

VisEvol$^1$$^4$ is a VA dashboard designed to facilitate hyperparameter search using evolutionary optimization (cf. Fig.~\ref{fig:system}(3)). By using multiple coordinated views, the dashboard enables users to create new hyperparameter sets and preserve robust ones within a majority-voting ensemble. Examining the effects of adding or removing algorithms and models in a majority-voting ensemble from various perspectives, along with monitoring the crossover and mutation processes, provides users with clarity on selecting hyperparameters for a single model or complex ensembles that require a balance of performance and diversity.

A concern for this dashboard is its reliance on projection-based views (e.g., produced by t-SNE).

\subsection{Model Understanding}

t-Distributed Stochastic Neighbor Embedding (t-SNE) is a widely used method for visualizing multidimensional data, with numerous successful applications across various fields. However, interpreting t-SNE projections can be challenging and, at times, misleading, which can undermine the reliability of the results. Gaining insight into the workings of t-SNE and the underlying causes of specific patterns in its output can be particularly difficult, especially for those unfamiliar with dimensionality reduction techniques.

The proposed workflow supports the selection of hyperparameters through visual exploration and quality metrics, offers a rapid assessment of projection accuracy, enables deeper investigation into quality by distinguishing the trustworthiness of various projection regions, and facilitates the interpretation of visible patterns in the projection relative to the original dataset's dimensions in an iterative process (see Fig.~\ref{fig:workflow}(4)).

t-viSNE$^1$$^5$ is a VA dashboard designed for exploring t-SNE projections visually (cf. Fig.~\ref{fig:system}(4)). By partially unveiling the inner workings of the t-SNE algorithm, we empower users to evaluate the quality of the projections and gain insights into the algorithm's decision making process when forming clusters. We also reveal typically overlooked information from within the algorithm, like point densities, and highlight regions that are suboptimally optimized by t-SNE.

With the foundation for assessing visual embeddings established, we now address understanding, debugging, and comparing supervised AI algorithms.

A simple, model-agnostic approach for XAI models involves training surrogate models, such as decision trees and rule sets, that approximate the original models while remaining simpler and more explainable. However, rule sets can become excessively lengthy with numerous if-else statements, and decision trees may quickly grow deep when attempting to replicate complex AI models accurately. Consequently, both methods can struggle to achieve their primary objective of delivering model interpretability to users.

The proposed workflow allows users to select a surrogate model that aligns with their preferred trade-off between fidelity and simplicity, examine decisions derived from the surrogate model (acting as a simplified version of the complex AI model), and adjust individual rules based on visual feedback and personal experience (see Fig.~\ref{fig:workflow}(5)). To complete the cycle, users can iteratively reason through specific test cases while reviewing previously generated explanations for the training data. 

DeforestVis$^1$$^6$ is a VA dashboard designed to employ decision stumps (single-level decision trees) derived from the AdaBoost algorithm, which are inherently interpretable, to create surrogate models that approximate the behavior of a target model (see Fig.~\ref{fig:system}(5)). The dashboard's linked views enable users to iteratively balance complexity and fidelity while reducing the precision (measured in decimal places) used in decision thresholds without compromising accuracy. Moreover, DeforestVis allows users to explore decision stumps in various ways, such as evaluating their purity and impact, and it summarizes the surrogate model's behavior by aggregating the predictive outcomes and influence of all decision stumps associated with each data feature. Users can modify rules and compare them at both local and global levels.

The following subsection explores the possible advantages of incorporating human involvement in debugging ensembles.

\subsection{Model Debugging}

Bagging and boosting are widely used ensemble techniques in AI that generate multiple individual decision trees. Because of their ensemble nature, these methods often achieve better predictive performance compared to single decision trees or other AI models. However, each decision tree creates numerous decision paths, increasing the model's complexity and limiting its application in fields requiring reliable and explainable outcomes. Therefore, as the number of decisions increases, the interpretability of algorithms such as Random Forest and AdaBoost diminishes.

The proposed workflow enables AI experts to identify high-performing and diverse models, prioritize important features, and refine models by adjusting hyperparameters (see Fig.~\ref{fig:workflow}(6)). Meanwhile, domain experts can investigate reliable decisions, benchmark them against global standards, identify local decisions for a specific test instance, and extract these insights.

VisRuler$^1$$^7$ is a VA dashboard that enables users to examine a range of rules derived from bagged and boosted decision trees in order to reach a consensus on the final decision for each case (cf. Fig.~\ref{fig:system}(6)). The dashboard’s multiple coordinated views support the selection of various high-performance models, analysis of feature contributions, management of multiple decisions, exploration of global decisions, and facilitation of case-based reasoning.

Having set the basis for combining bagging and boosting methods in various voting schemes, the next subsection explores the potential benefits of involving humans in the construction of stacking ensembles.

\subsection{Model Comparison}

Stacking is an ensemble method that integrates various base models, organized in at least one layer, and uses a metamodel to aggregate their predictions. While it can significantly improve the predictive accuracy of AI, creating a model stack from the ground up often involves a time-consuming trial-and-error process. This difficulty arises due to the vast range of possible solutions, including varying sets of data instances and features for training, a wide selection of algorithms, and different parameter settings (i.e., models) that yield different performances based on various metrics.

The proposed workflow is built on a knowledge generation model for ensemble learning using VA (see Fig.~\ref{fig:workflow}(7)). On the left, it shows how a VA dashboard can facilitate data and model exploration through visualization. On the right, several design objectives support the user in exploring, verifying, and generating knowledge for ensemble learning.

StackGenVis$^1$$^8$ is a VA dashboard designed for aligning data, algorithms, and models in stacking ensemble learning. Through multiple coordinated views, users can create an effective stacking ensemble from scratch. By exploring the algorithms, data, and models from various perspectives and monitoring the training process, users can confidently determine the next steps in developing complex stacks of models, requiring a combination of both the highest-performing and the most diverse individual AI models.

\begin{figure*}
\centerline{\includegraphics[width=37pc]{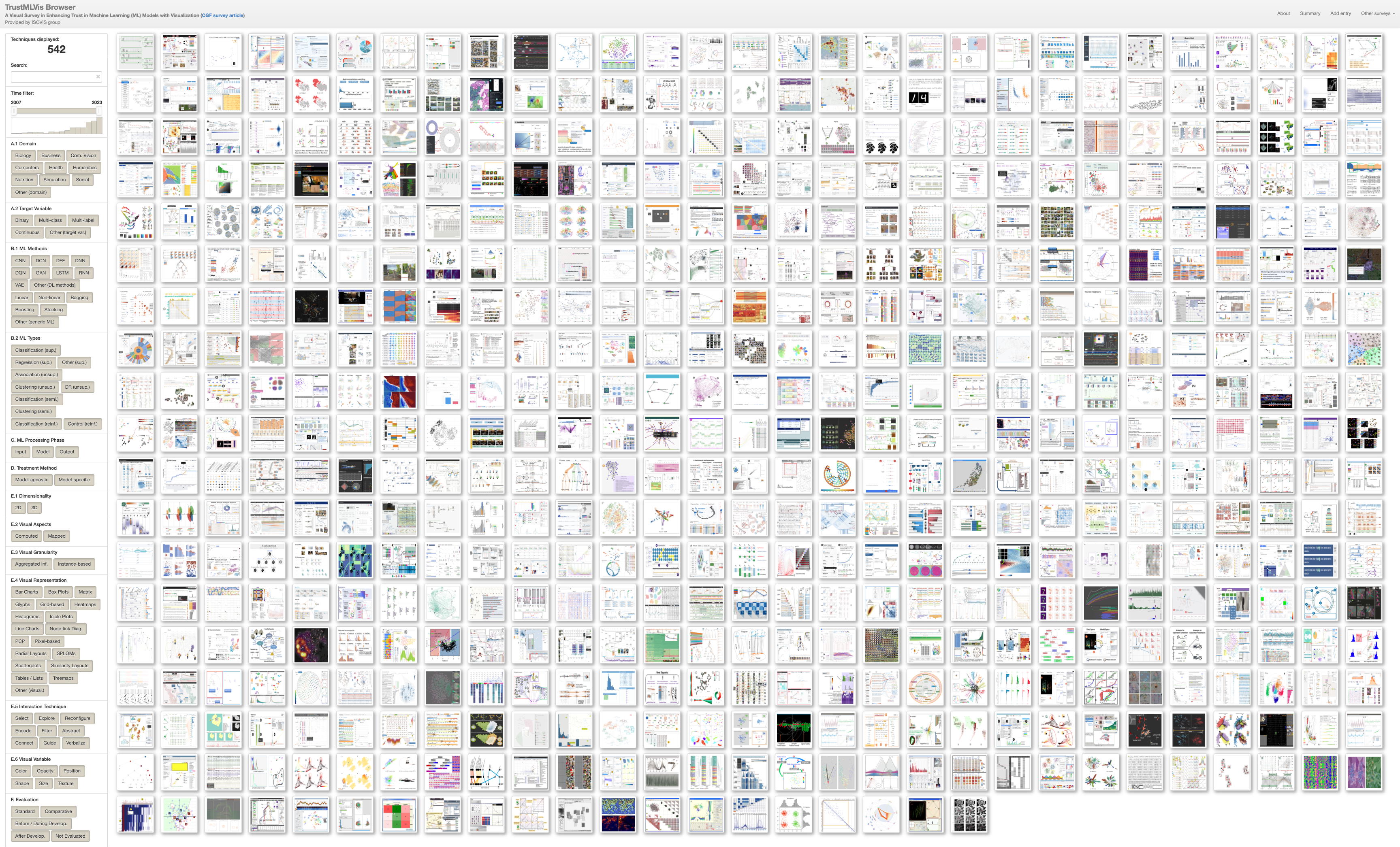}} \vspace{-2mm}
\caption{Our interactive survey browser$^9$$^,$$^1$$^0$ for exploring the 542 identified VA techniques (available at \href{https://trustmlvis.lnu.se}{trustmlvis.lnu.se}).}
\label{fig:web} \vspace{-1mm}
\end{figure*}

\section{LESSONS LEARNED}
\subsection{Ensemble Learning for Tabular Data}
The VA community has devoted significant attention to XAI in recent years, yet much of the focus remains on NNs for image data. Out of 542 visualization papers surveyed on enhancing trust in AI, 345 primarily addressed NN models, while only 114 concentrated on ensembles. Although image analysis is undeniably an important research domain---especially in fields like medicine, where NNs excel with high-dimensional data---other application areas should not be overlooked. Tabular data still represents the majority of datasets employed in AI workflows, where ensemble learning models often surpass NNs in performance.$^7$

\subsection{Ideas Beyond Developing VA Dashboards} 
The visualization approaches discussed in this work can also be employed as standalone tools to support various stakeholders (refer to Table~\ref{tab:categorization}) in addressing complex challenges related to different phases of the AI pipeline (cf. Fig.~\ref{fig:outline}). However, the individual VA workflows proposed are highly adaptable and could be integrated to create custom solutions that address increasingly difficult supervised and unsupervised learning tasks. This integration enables the creation of a versatile design space with personalized approaches for managing AI models designed for tabular data, explored from multiple perspectives, ranging from local explanations of single instances to global summaries of model behavior across all or specific subsets of data.

\subsection{Common Sources of Bias} 
In this work, the assumption of ground truth in supervised classification tasks was maintained across all use cases and usage scenarios. However, evaluating the quality and fairness of the data, as well as identifying potential biases or errors from the annotators responsible for labeling, could prove critical. Also, visualizations have the potential to obscure additional biases. For example, in VisEvol, we developed a bar chart that represents the uncertainty introduced by the force-based layout algorithm used for the beeswarm plot, as shown in Fig.~\ref{fig:system}(3). Additionally, incorporating domain knowledge into visual explanation dashboards can introduce bias, as data scientists might fine-tune parameters to align explanations with their own assumptions.$^1$$^9$ This issue is not exclusive to our approach but is inherent in the application of domain-specific expertise. Hence, it is essential to exercise caution and explicitly signal instances where explanations diverge from the actual model by using provenance and uncertainty visualizations.

\subsection{Online Survey Browser} 
Our publicly accessible survey browser has received considerable positive feedback from the VA community for facilitating the discovery of new papers. Our web-based survey browser shown in Fig.~\ref{fig:web} has been visited by nearly 5,000 unique users from 122 countries since its launch in Spring 2020 (as reported by Google Analytics). We consider TrustMLVis Browser a valuable resource across multiple disciplines and remain committed to maintaining and further developing it in the future. For example, the initial dataset of 200 entries has recently grown to 542, as shown in Table~\ref{tab:categorization}.

\subsection{Teaching Applications} 
Several VA dashboards developed by us have demonstrated their potential in teaching fundamental AI concepts to computer science students who are comfortable with visualizations. For two consecutive years, students in the advanced InfoVis course at Linnaeus University (Sweden) were tasked with exploring Google’s Embedding Projector and providing feedback on its key strengths, weaknesses, and potential improvements. Next, t-viSNE was introduced, enabling students to compare the two dashboards in addressing the shortcomings of t-SNE highlighted by Wattenberg et al.$^2$$^0$ During the t-viSNE paper's presentation at the IEEE VIS 2020 conference, attendees noted the clear t-viSNE's applicability in educational settings. Also, an AI expert evaluating VisEvol suggested that features like the Sankey diagram could assist students in grasping the concept of evolutionary optimization. Another AI expert acknowledged the educational value of HardVis.

\section{IMPACTS AND FUTURE WORK}
Under the General Data Protection Regulation (GDPR) and the AI Act, data controllers must provide clear explanations of the logic behind automated decisions. Additionally, individuals have the right to avoid being subjected to decisions made solely through automated processes. Enhancing trust in human decision making influenced by AI algorithms is a crucial step in addressing concerns about such outcomes, particularly when aiming to minimize biases. For example, in healthcare, medical professionals and data analysts who use AI models to diagnose diseases based on sensitive medical data bear both ethical and legal responsibility for the resulting decisions. In modern practice, failing to explain how a specific medical prognosis was reached is increasingly viewed as unacceptable. As the volume of digital data continues to expand and AI technology plays a growing role in societal functions, computational and VA methods can offer powerful tools for addressing core challenges. This work demonstrates how VA dashboards can foster greater explainability and trust in high-stakes decision making like healthcare, where people lives are at stake.

The methods described in this work are applicable to both academic research and business intelligence. For example, in the financial sector, declining a loan application should involve greater transparency, clearly outlining the reasons for rejection. Building trust between humans and algorithms depends on understanding how the algorithm operates and being able to explain the reasoning behind its predictions.

Finally, the research opportunities (R1--R8) identified in our 2024 STAR$^1$$^0$ offer valuable insights for visualization researchers, practitioners across diverse disciplines, and AI experts. We summarize them below.

\subsection{R1: Improving Popular XAI Methods}
Many widely recognized XAI methods using visualization to communicate results were developed by experts from disciplines outside of visualization research. The growing appeal of these model-agnostic techniques, such as LIME and SHAP, stems from their perspective of a model as a mathematical function. These functions can be broken down into simpler components, whose behavior can then be evaluated using interpretation methods. Visualization researchers can enhance these methods by designing VA dashboards that improve their interactivity and visual representations or by directly refining the algorithmic components of the techniques with their specialized knowledge.

\subsection{R2: New NN Approaches \& Self-Supervision}
The AI community consistently develops new NN architectures or adapts existing ones to different contexts, whereas visualization researchers have been slower in adopting and exploring these advancements. For example, transformer NNs have shown promising results in computer vision tasks compared to convolutional NNs, yet the VA community has rarely engaged in visually supporting their exploration. Also, self-supervision represents a rapidly growing field that warrants greater attention from the VA community, alongside prompt engineering and model checkers to address ``hallucinations'' in large language models (LLMs).

\subsection{R3: Multiverse \& Confirmatory VA}
Recently, most VA solutions have predominantly emphasized exploratory visual analysis. However, there is a growing trend toward employing visual representations for confirmatory analysis, hypothesis testing, and causal reasoning (e.g., counterfactual reasoning, causal learning, and causal inference), which could benefit from enhanced visualization. Another relevant question is how VA dashboards can assist users in conducting a comprehensive multiverse analysis to identify optimal model strategies, enabling them to understand complex models better and make informed decisions about a specific course of action.

\subsection{R4: Input/Output Uncertainty Quantification}
Most VA dashboards are designed to assist users in understanding models, but two significant challenges remain: quantifying input and output uncertainty and evaluating robustness and sensitivity to changes, as users aim to deploy dynamic, well-calibrated models in practical real-world scenarios. While uncertainty quantification and sensitivity analysis are widely used in other disciplines, visualization research still lacks comprehensive approaches for examining model inputs and outputs, such as the conformal prediction method, which offers statistical guarantees.

\subsection{R5: Rigorous Evaluation \& Benchmarking} 
Collaboration with domain experts from various disciplines often encounters significant communication challenges. These stem from differences in terminology, expectations, and expertise, requiring considerable effort to align goals and solutions. To reduce reliance on expensive user evaluations, it is crucial to explore AI-assisted, simulated evaluation methods. Enhancing the rigor of state-of-the-art benchmarking datasets is equally important, as they may contain notable errors. Additionally, the development of rational agent benchmarks that assess the necessity and advantages of visualization can support the evaluation of XAI methods. Such benchmarks can potentially reduce the dependence on extensive human involvement, thereby lowering associated costs.

\subsection{R6: Model Deployment \& Visual Channels}
Scaling VA dashboards to handle a large number of instances, features, and algorithms is particularly challenging, especially in multi-class classification contexts. While progressive VA combined with incremental learning offers a promising approach, developing simpler yet more robust systems is critical for real-world applications, where issues like covariate and label shifts frequently occur. The simultaneous use of visual channels, such as color and opacity, for encoding tasks adds to the complexity. Many users struggle to navigate these sophisticated dashboards, particularly during their initial interactions. Visualization literacy can help address this issue by educating users. In addition to existing tutorials, storytelling can effectively be used to demonstrate a dashboard's functionalities, while sonification and verbalization can enhance user comprehension and simplify the use of VA dashboards.

\subsection{R7: Advancing the Impact \& Reproducibility}
Many application-specific VA dashboards fail to transition into real-world deployment, leaving them vulnerable to concept and distribution drifts. This underscores the necessity of investigating the applicability of models in out-of-domain/distribution scenarios, stressing the importance of adapting specialized models into more flexible AI solutions. Broadening the accessibility and usability of VA dashboards can be achieved through various approaches, such as incorporating visualizations within computational notebooks and providing hybrid solutions that combine programmatic APIs with interactive UIs to gather user feedback over extended periods. Adopting open-source practices plays a major role in encouraging community participation and addressing key challenges related to the replication and reproducibility of AI models used in VA dashboards.

\subsection{R8: Underexplored Areas}
By analyzing the categorization of the 542 visualization techniques, it becomes evident that the underrepresented categories have the potential to inspire new research directions for non-TL areas, indirectly impacting trust in AI. For example, VA dashboards for boosting and stacking ensemble learning methods are notably scarce. Similarly, multi-label and regression problems are less explored compared to classification. In the realm of unsupervised learning, association and pattern mining remain underrepresented in VA. Reinforcement learning also receives minimal attention. Lastly, more emphasis is needed on addressing the needs of beginners and users with varying levels of experience who aim to analyze their data broadly.

\section{ACKNOWLEDGMENTS}

The author would like to thank his main Ph.D. advisor, Prof. Dr. Andreas Kerren (Linköping University, Sweden), for the tremendous support and guidance for this dissertation work. Also, many thanks to his Ph.D. co-advisors, Dr. Rafael M. Martins (Linnaeus University, Sweden) and Dr. Ilir Jusufi (Blekinge Institute of Technology, Sweden), for the feedback received throughout the entire Ph.D. study. The author is thankful to his postdoc mentor, Prof. Dr. Jessica Hullman (Northwestern University, USA), for her efforts in teaching me her perspective on conducting research. Finally, many thanks to Prof. Dr. Alexandru C. Telea for his mentorship in further advancing my academic career.

\def\refname{REFERENCES}

\vspace*{-8pt}

\begin{IEEEbiography}{Angelos Chatzimparmpas}{\,} is an Assistant Professor with the Department of Information and Computing Sciences at Utrecht University, Utrecht, 3584 CC, The Netherlands. His research interests include visual analytics, human-computer interaction, machine learning interpretability, human-centered AI, and XAI systems. Chatzimparmpas received his Ph.D. degree in computer and information science from Linnaeus University, Växjö, Sweden, in 2023. He is a member of IEEE and the corresponding author of this article. Contact him at a.chatzimparmpas@uu.nl.
\end{IEEEbiography}

\hfill \break \break \vspace*{-2pt}
Contact department editor Sumanta N. Pattanaik at Sumanta.Pattanaik@ucf.edu.

\end{document}